\documentclass[%
 reprint,%
 amssymb, amsmath,api,cha,aps,prb%
]{revtex4-1}
\usepackage{mhchem}
\usepackage{siunitx}
\usepackage{graphicx}
\usepackage{color}
\usepackage{docs}%
\usepackage{bm}%
\usepackage[colorlinks=false,linkcolor=blue]{hyperref}%
\expandafter\ifx\csname package@font\endcsname\relax\else
 \expandafter\expandafter
 \expandafter\usepackage
 \expandafter\expandafter
 \expandafter{\csname package@font\endcsname}%
\fi
\def\newslash{\slash}
\def\newdash{-}
\newif\ifmoreelements
\newif\ifmorelayers

\def\Ce#1{%
    \def\templayer{#1}%
    \loop%
        \expandafter\Ceii\templayer-=%
        \expandafter\ce\expandafter{\tempfirstelement}%
    \ifmoreelements%
        \newdash%
        \expandafter\Cei\tempotherelements=%
    \repeat}
\def\Cei#1-={%
    \def\templayer{#1}}
\def\Ceii#1-#2={%
    \def\tempfirstelement{#1}\def\tempotherelements{#2}%
    \ifx\tempotherelements\empty\moreelementsfalse\else\moreelementstrue\fi}
\def\CE#1{%
    \def\tempstack{#1}%
    \loop%
        \expandafter\CEii\tempstack/+%
        {\expandafter\Ce\expandafter{\tempfirstlayer}}%
    \ifmorelayers%
        \newslash%
        \expandafter\CEi\tempotherlayers+%
    \repeat}
\def\CEi#1/+{%
    \def\tempstack{#1}}
\def\CEii#1/#2+{%
    \def\tempfirstlayer{#1}\def\tempotherlayers{#2}%
    \ifx\tempotherlayers\empty\morelayersfalse\else\morelayerstrue\fi}
\begin{document}

\title{Titanium Nitride as a Seed Layer for Heusler Compounds}%

\author{Alessia Niesen}%
\email{aniesen@physik.uni-bielefeld.de}
\affiliation{Center for Spinelectronic Materials and Devices, Bielefeld University, Germany}%
\author{Manuel Glas}
\affiliation{Center for Spinelectronic Materials and Devices, Bielefeld University, Germany}%
\author{Jana Ludwig}
\affiliation{Center for Spinelectronic Materials and Devices, Bielefeld University, Germany}%
\author{Roshnee Sahoo}
\affiliation{Max Planck Institute for Chemical Physics of Solids, Dresden, Germany}%
\author{Daniel Ebke}
\affiliation{Max Planck Institute for Chemical Physics of Solids, Dresden, Germany}%
\author{Elke Arenholz}
\affiliation{Advanced Light Source, Lawrence Berkeley National Laboratory, Berkeley, California 94720, USA}
\author{Jan-Michael Schmalhorst}
\affiliation{Center for Spinelectronic Materials and Devices, Bielefeld University, Germany}%
\author{G\"unter Reiss}
\affiliation{Center for Spinelectronic Materials and Devices, Bielefeld University, Germany}%

\date{\today}%

\begin{abstract}
Titanium nitride (\ce{TiN}) shows low resistivity at room temperature (\SI{27}{\micro\ohm \cm}), high thermal stability and thus has the potential to serve as seed layer in magnetic tunnel junctions. High quality \ce{TiN} thin films with regard to the crystallographic and electrical properties were grown and characterized by x-ray diffraction and 4-terminal transport measurements. Element specific x-ray absorption spectroscopy revealed pure \ce{TiN} inside the thin films. To investigate the influence of a \ce{TiN} seed layer on a ferro(i)magnetic bottom electrode in magnetic tunnel junctions, an out-of-plane magnetized \ce{Mn_{2.45}Ga} as well as in- and out-of-plane magnetized \ce{Co2FeAl} thin films were deposited on a \ce{TiN} buffer, respectively. The magnetic properties were investigated using a superconducting quantum interference device (SQUID) and anomalous Hall effect (AHE) for \ce{Mn_{2.45}Ga}. Magneto optical Kerr effect (MOKE) measurements were carried out to investigate the magnetic properties of \ce{Co2FeAl}. \ce{TiN} buffered \ce{Mn_{2.45}Ga} thin films showed higher coercivity and squareness ratio compared to unbuffered samples. The Heusler compound \ce{Co2FeAl} showed already good crystallinity when grown at room temperature on a \ce{TiN} seed-layer.
\end{abstract}

\maketitle

\section{Introduction} 
Spintronic exploits the influence of the electron's spin on its transport in solids. The key component of spintronic applications, like magnetic sensors or storage media, is the magnetic tunnel junction (MTJ). Spin dependent transport phenomena in MTJ's, called tunnel magnetoresistance (TMR) effect, could be maximized for materials with high spin polarization. Thus half-metallic materials are preferred, due to a spin polarization of \SI{100}{\%} at the Fermi level. The half-metallic characteristic has been predicted for oxide compounds like \ce{Fe3O4} (Magnetite) and several Heusler compounds.\cite{Versluijs:2001uf,DEGROOT:1983vn} To achieve high crystalline ordering of the thin films, the lattice mismatch between the material and the substrate or seed layer has to be minimized. In addition a metallic buffer layer could act as a conduction layer. Common seed layers for Heusler compounds are \ce{Cr} (\(a_{\rm{Cr}}\) = \SI{2.88}{\angstrom}) and \ce{Pt} (\(a_{\rm{Pt}}\) =  \SI{3.92}{\angstrom}). Co-based Heusler compounds, like \ce{Co2FeAl} (\ce{CFA}), \ce{Co2FeSi} (\ce{CFS}), or \ce{Co2MnSi} (\ce{CMS}) have a lattice constant of \(a_{\rm{Heusler}} \approx \SI{5.7}{\angstrom}\). Therefore the use of a \ce{Cr} buffer or \ce{MgO} substrate leads to epitaxial growth.\cite{Sterwerf:2013it} To maintain the thermal stability at shrinking device sizes, an out-of-plane oriented magnetization of the material is advantageous. Therefore perpendicularly magnetized \ce{Mn_{3-x}Ga} (\(0.15 \leq x \leq 2\)) compound found recently a lot of interest. 
The perpendicular magnetocrystalline anisotropy (PMA) of the binary \ce{Mn}-\ce{Ga} compound is an intrinsic effect based on a crystal anisotropy. The spin polarization at the Fermi level of \ce{Mn}-\ce{Ga} is predicted to be \SI{88}{\%} for the tetragonally distorted phase.\cite{Balke:2007eb} This crystalographic phase is formed if the mismatch between substrate and the thin film is small. In addition, the transition from the cubic \(\mathrm{D0_{3}}\) into the tetragonal \(\mathrm{D0_{22}}\) phase takes place at temperatures above 500\si{\degreeCelsius}.\cite{Glas:2013we} To increase the applicability of these materials in magnetic tunnel junctions an optimized buffer layer is needed. \ce{Pt} is a promising material, due to the in-plane lattice constant of 3.92\,\AA\ and a lattice mismatch of 0.2\,\% to the in-plane lattice constant of the tetragonally distorted \ce{Mn}-\ce{Ga} (\(\mathrm{D0_{22}}\)). Unfortunatly the favored growth direction of Pt on MgO (100) and \ce{SrTiO3} (100) substrates is the (111) direction. Crystalline growth in the (001) direction is achieved by complicated preparation techniques. Furthermore \ce{Pt} is a critical raw material compared to Cr or Ti. For 45\,degree rotated growth of the \ce{Mn}-\ce{Ga} the lattice mismatch with \ce{Cr} is 5\,\%. Several groups reported crystalline \ce{Mn}-\ce{Ga} thin films on \ce{Cr} buffered \ce{MgO} substrates. The diffusion temperature of Cr, however, is around 450\si{\degreeCelsius}, while \ce{Mn}-\ce{Ga} requires a deposition temperature of around 550\si{\degreeCelsius}. The diffusion problem also lowers the applicability of a Cr buffer for the Co-based Heusler compounds, like \ce{Co2FeAl}, Co\(_2\)FeSi and Co\(_2\)MnSi. To achieve high crystalline order of these compounds (L2\(_1\) crystal structure) post-annealing processes have to be carried out at temperatures around \SI{500}{\degreeCelsius}. Investigations showed decreased TMR after the post annealing at such high temperatures, explained by \ce{Cr} diffusion into the ferromagnetic electrode.\cite{Ebke:2010ci,Sterwerf:2013it} 
Due to the high thermal stability (melting point \SI{2950}{\degreeCelsius})\cite{Pritschow:07} interdiffusion of \ce{TiN} is prevented. Another advantage is the low electrical resistivity of sputter deposited TiN (\SI{16}{\micro\ohm \cm}) and a surface roughness below \SI{1}{\nano\m}.\cite{Magnus:2011wl,Krockenberger:2012bo} 
The lattice constant of \ce{TiN} (fcc structure) is \SI{4.24}{\angstrom} and therefore suitable for various Heusler compounds. By rotating the unit cell of the Co-based Heusler compounds by 45\,degree a lattice mismatch of about \SI{5}{\%} is achieved. 
\newline
We investigated the structural properties of sputter deposited \ce{TiN} on single crystalline \ce{MgO} (001) and \ce{SrTiO3} (001) substrates at different deposition temperatures using x-ray diffraction (XRD) and reflection (XRR). Since the surface properties are of large significancy for applications, atomic force microscopy (AFM) was carried out to verify the surface roughness. Temperature dependent transport measurements were realized in a closed cycled helium cryostat, with a temperature range from \SIrange{2}{300}{\kelvin}. X-ray absorption spectroscopy (XAS) was performed at beamline (BL) 6.3.1 of the Advanced Light Source in Berkeley. The Nitrogen \(K\)-edge and Titanium \(L\)-edges were investigated by surface sensitive total electron yield (TEY) and bulk sensitive luminescence mode (LM) in normal incidence. Furthermore, we investigated the crystallographic and magnetic properties of \ce{Co2FeAl} and \ce{Mn_{3-x}Ga} with \ce{TiN} seed layer deposited on \ce{MgO} and \ce{SrTiO3} substrates.

\begin{figure}[t!]%
\includegraphics[width=\linewidth]{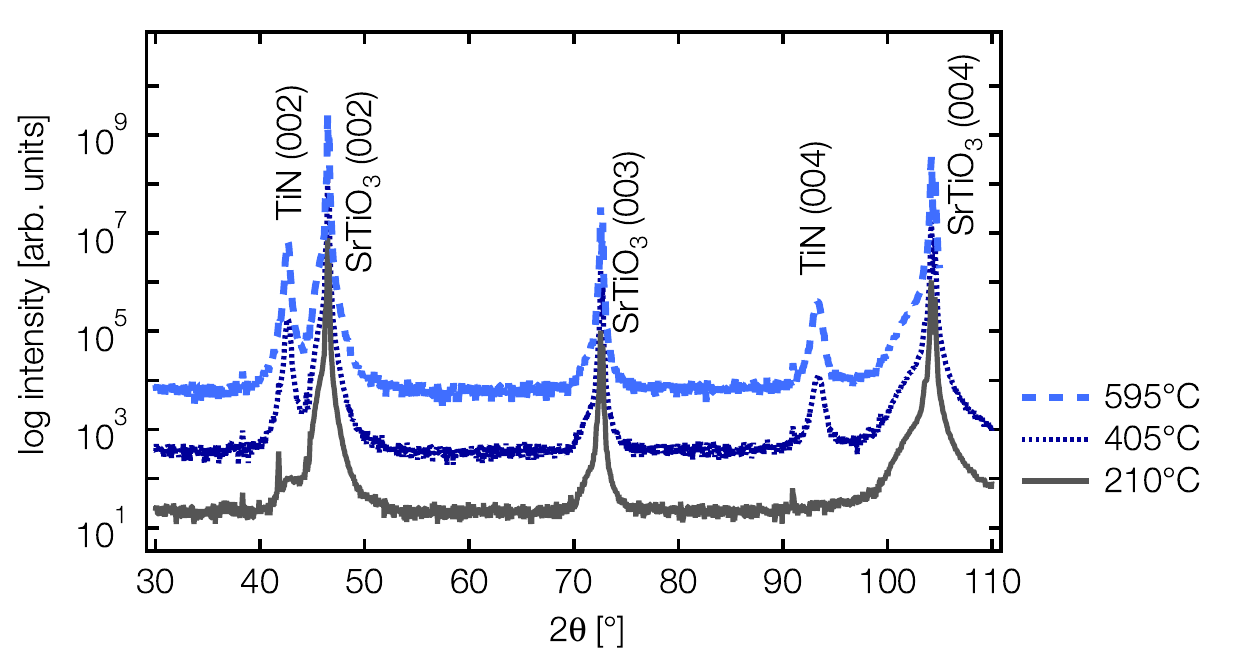}%
\caption{X-ray diffraction pattern for DC sputter deposited \ce{TiN} thin films (\SI{30}{\nano\m}) on \ce{SrTiO3} substrates at \SIlist{210;405;595}{\degreeCelsius}.}%
\label{fig:TiN_STO}%
\end{figure}%
\begin{figure}[t!]%
\includegraphics[width=3in]{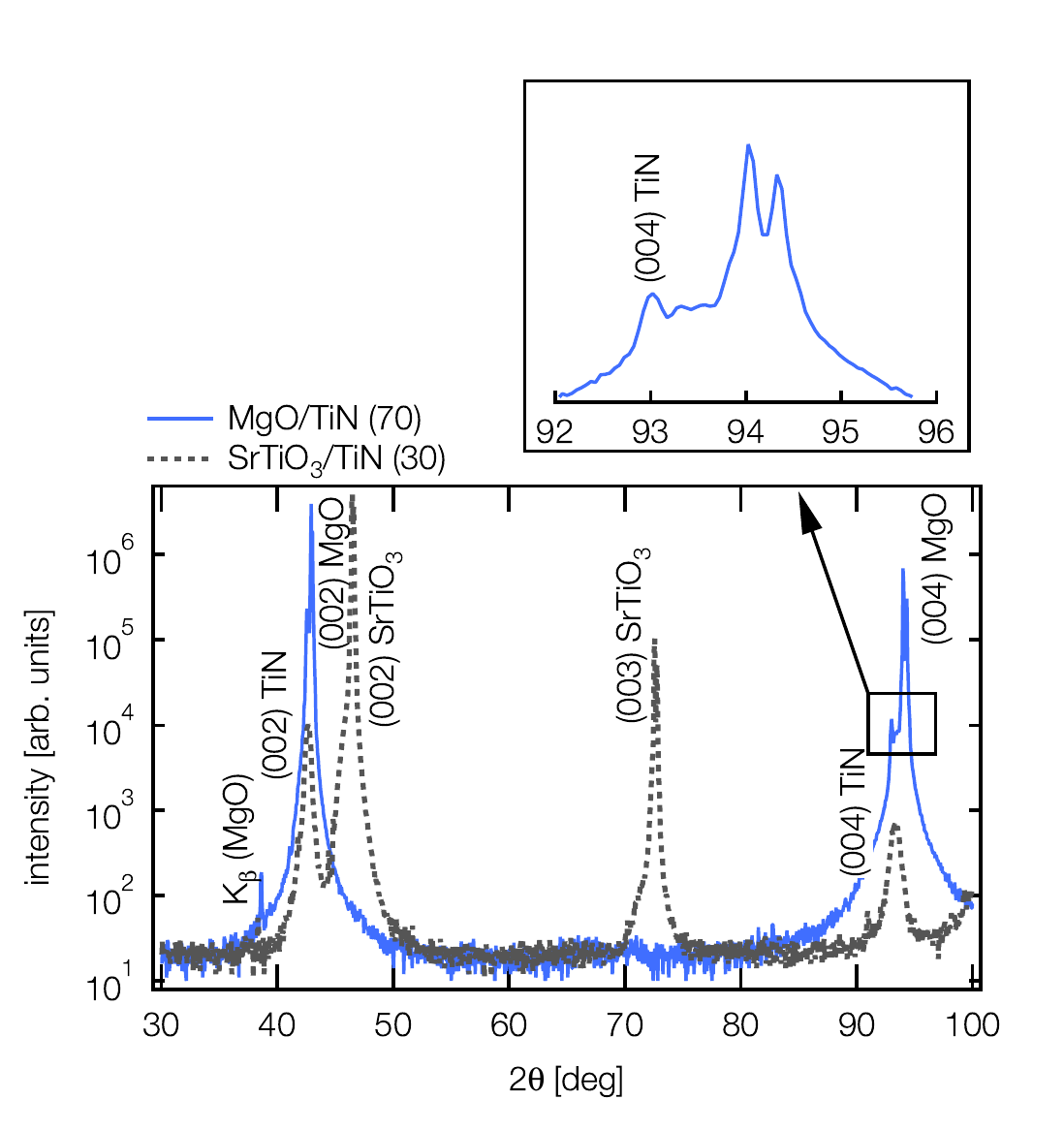}%
\caption{X-ray diffraction pattern for \ce{TiN} thin films (thickness is given in parenthesis) on \ce{MgO} (blue) and \ce{SrTiO3} (grey dashed) substrate. The cutout on top shows Laue oscillations around the (004) \ce{TiN} peak.}%
\label{fig:TiN_laue}%
\end{figure}%

\section{Experimental}

DC and RF magnetron sputtering was used to prepare the samples. The \ce{TiN} as well as the \ce{Mn_{2.45}Ga} layers were deposited in an UHV sputtering system with a base pressure below \SI{3e-9}{\milli\bar}. Reactive sputtering in an Argon - Nitrogen atmosphere results in stoichiometric \ce{Ti1N1} thin films. During the sputtering a \ce{N} flow of \SI{2}{sccm} combined with \SI{20}{sccm} \ce{Ar} was used, leading to a deposition pressure of \SI{1.6e-3}{\milli\bar}. Because of the low lattice mismatch with \ce{TiN} (below \SI{0.7}{\%}) \ce{MgO} (100) single crystalline substrates were used. \ce{SrTiO3} (100) substrates were utilized, due to the low mismatch with the \ce{Mn}-\ce{Ga} crystal (below \SI{1}{\%}). 
A \ce{Mn60Ga40} composite target and an Ar pressure of \SI{1.7e-3}{\milli\bar} were used to deposit the \ce{Mn_{2.45}Ga} thin films on top of the TiN. The deposition temperature \(T_{dep}\) of \ce{TiN} was \SIlist{210;405;595;830}{\degreeCelsius}, respectively. Whereas the \ce{Mn}-\ce{Ga} was deposited at \SIlist{550;595}{\degreeCelsius}. In addition \ce{TiN} buffered \ce{Co2FeAl} thin films on \ce{MgO} and \ce{SrTiO3} substrates were prepared. For the \ce{Co2FeAl}, the substrates with a \ce{TiN} buffer layer were transfered to another magnetron sputtering machine, without vacuum break. Here the base pressure was \SI{1e-7}{\milli\bar}. Stoichiometric \ce{Co2FeAl} was deposited from a composite target under an \ce{Ar} pressure of \SI{2.3e-3}{\milli\bar}. On top of the Heusler compound a \SI{2}{\nano\m} thick \ce{MgO} layer was deposited to protect the stack from degradation.

\section{Results}

\subsection{Chrystallographic, structural and electrical properties of the TiN seed-layer}

Crystallographic properties of the \ce{TiN} thin films (\SI{30}{\nano\m}) were determined using an X'Pert Pro diffractometer (\ce{Cu} anode). The \ce{TiN} films showed no dependence of their crystalline quality on the deposition temperature on MgO substrates. Even at \SI{210}{\degreeCelsius} an epitaxial growth was obtained (not shown). On the contrary, \ce{TiN} layers on \ce{SrTiO3} required a deposition temperature higher than \SI{210}{\degreeCelsius} to achieve epitaxial growth (Figure \ref{fig:TiN_STO}). Figure \ref{fig:TiN_laue} shows an XRD pattern comparison of \ce{TiN} deposited on \ce{MgO} and \ce{SrTiO3} substrate at \SI{405}{\degreeCelsius}. The \SI{70}{\nano\m} thick \ce{TiN} film deposited on \ce{MgO} shows Laue oscillations, which are a clear evidence for high crystalline coherence (cutout in Figure \ref{fig:TiN_laue}). Due to the low lattice mismatch with \ce{MgO} the \ce{TiN} film reflexes are close to the substrate peaks and therefore difficult to investigate. 
The out-of-plane lattice constant, determined by the shoulder next to the (002) and (004) \ce{MgO} reflex, is \(c = 4.25\)\,\AA. On \ce{SrTiO3} substrates, the reflexes of the TiN layer are clearly visible and exhibit the same out-of-plane lattice constant.
\begin{figure}[t!]%
\vspace{-0.3cm}
\hspace{-0.5cm}
\includegraphics[width=\linewidth]{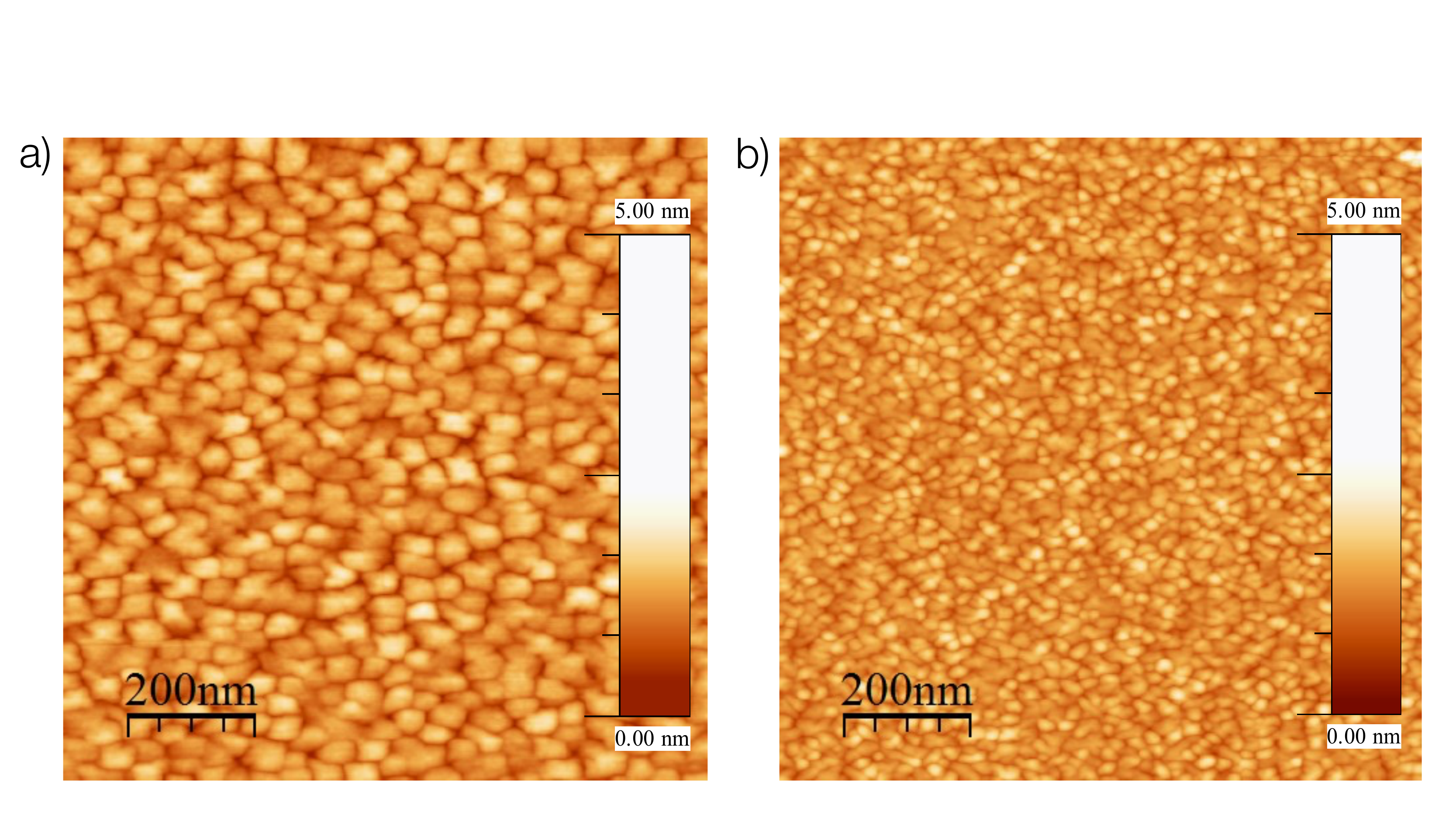}%
\caption{Atomic force microscopy images ($\SI{1}{\micro\m^2}$ section) of \ce{TiN} thin films deposited at \SI{595}{\degreeCelsius} on a) \ce{MgO}  and b) \ce{SrTiO3}  substrates. The grain size on \ce{SrTiO3} substrate is \(25\pm5\)\,\SI{}{\nano\m} compared to \(50\pm5\)\,\SI{}{\nano\m} on \ce{MgO}.}%
\label{fig:AFM_TiN}%
\end{figure}%
\begin{figure}[t!]%
\includegraphics[width=3in]{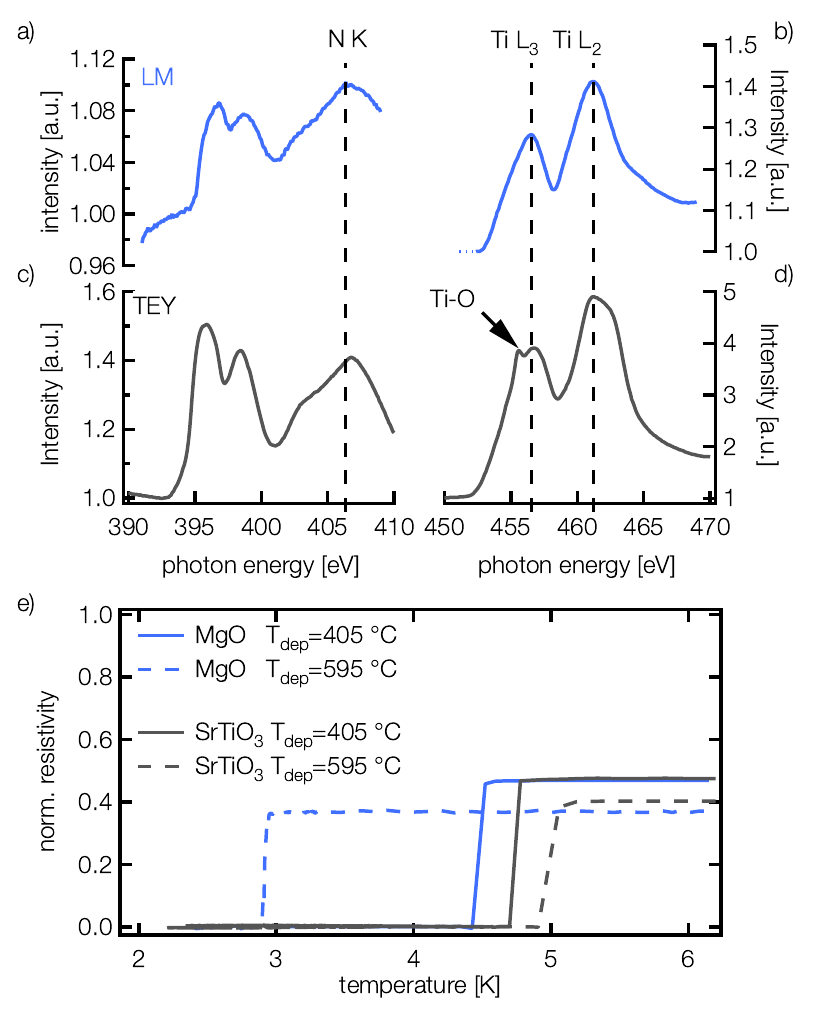}%
\caption{X-ray absorption spectra of a) and c) \ce{N} and b) and d) \ce{Ti} taken in total electron yield (grey) and luminescence mode (blue). The sample was deposited at \SI{405}{\degreeCelsius} on \ce{MgO} substrate. e) Temperature dependence of the resistivity down to \SI{2}{\kelvin} for \ce{TiN} thin films deposited at \SIlist{405;595}{\degreeCelsius}.}%
\label{fig:XAS_TiN_N_Ti}%
\end{figure}%
\begin{figure}[t!]%
\includegraphics[width=3in]{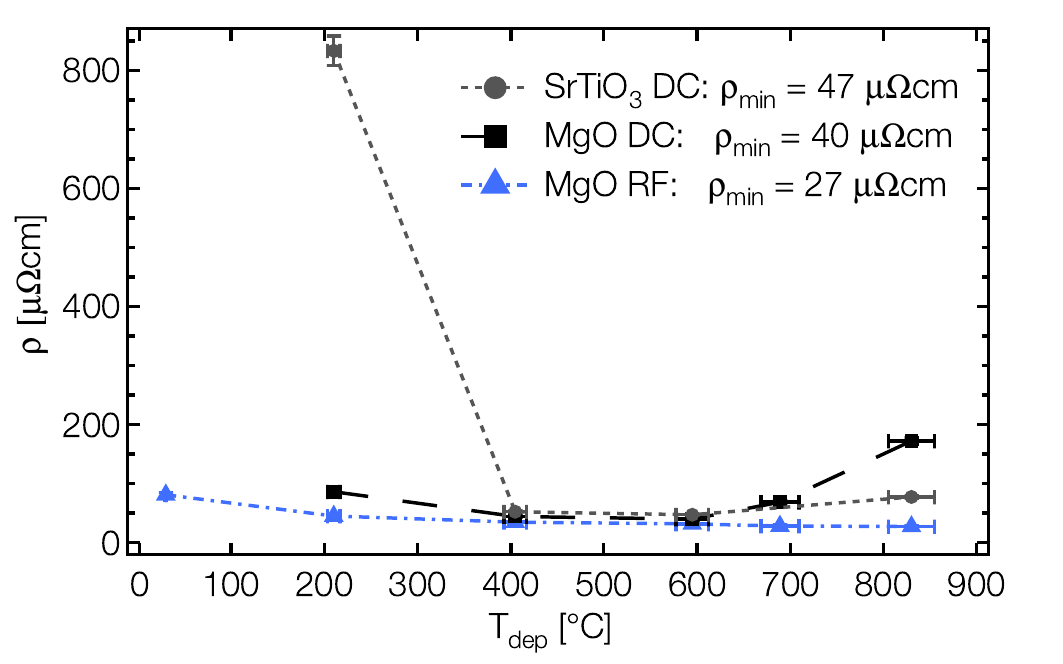}%
\caption{Resistivity dependence (measured at room temperature) of TiN on the deposition temperature \(T_{dep}\) and the substrate. The high resistivity value for \ce{TiN} deposited on \ce{SrTiO3} at \SI{210}{\degreeCelsius} confirms that there is no crystalline growth.}%
\label{fig:resist_TiN}%
\end{figure}%
Atomic force microscopy and x-ray reflection measurements were carried out to investigate the surface roughness of the \ce{TiN} layers. The roughness obtained by XRR is between \SIlist{0.6;1}{\nano\m} for all samples. The lowest value of \(0.6\pm0.1\) \SI{}{\nano\m} was obtained for DC deposited \ce{TiN} with a deposition temperature of \SI{595}{\degreeCelsius} on both substrate types. AFM measurements of this samples (on a \SI{3}{\micro\m^2} section) showed an RMS (root mean square) value of \SI{0.2}{\nano\m} on \ce{MgO} (Figure \ref{fig:AFM_TiN} a) and \SI{0.5}{\nano\m} on \ce{SrTiO3} (Figure \ref{fig:AFM_TiN} b) substrate. The lower roughness values determined by AFM are attributed to the smaller scanned section compared to the XRR measurements. Another reason is that the AFM is only sensitive to the surface, so that possible gaps between the grains could not be detected. Whereas x-rays are penetrating into the sample surface and therefore are sensitive to steep grain boundaries. However, on \ce{SrTiO3} both values are in good agreement, which was attributed to the smaller grain size of \(25\pm5\)\,\SI{}{\nano\m} compared to \(50\pm5\)\,\SI{}{\nano\m} on \ce{MgO}. Due to the reduced lateral grain size we obtain more grain boundaries on the scanned \SI{3}{\micro\m^2} section, which leads to increased roughness.\newline
The chemical properties of \ce{TiN} thin films were investigated using XAS. Figure \ref{fig:XAS_TiN_N_Ti} a) - d) depicts the X-ray absorption spectra of \ce{N} and \ce{Ti} taken in total electron yield TEY (grey) and luminescence mode LM (blue). The weak multiplet structure of the \(L_3\)-\ce{Ti}-edge indicates a small amount of \ce{Ti}-\ce{O} on the surface. However, the bulk sensitive LM spectra of \ce{Ti} and \ce{N} are in good agreement with the literature.\cite{Soriano:1993vd}

\begin{table}[!t]
\centering
\begin{tabular}{|c||c||c|}
\hline
\(T_{dep}\) (\SI{}{\degreeCelsius}) & on \ce{MgO} (100) & on \ce{STO} (100)\\
\hline
405 & \SI{4.65}{\kelvin} & \SI{4.78}{\kelvin}\\
\hline
595 & \SI{2.94}{\kelvin} & \SI{5}{\kelvin}\\
\hline
\end{tabular}
\caption{Dependence of the transition temperature into the superconducting state of sputter deposited \ce{TiN} on the substrate and the deposition temperature. }
\label{tab:Tc}
\end{table}
Temperature dependent transport measurements in a closed cycled \ce{He}-cryostat revealed a transition into the superconducting state. The transition temperature \(T_C\) showed a dependence on the substrate as well as on the deposition temperature of \ce{TiN} (Table \ref{tab:Tc} and Figure \ref{fig:XAS_TiN_N_Ti} e). 
The highest \(T_C\) of \SI{5}{\kelvin} was achieved for \ce{TiN} deposited at \SI{595}{\degreeCelsius} on \ce{SrTiO3} substrate. However, the same stack on \ce{MgO} revealed the lowest \(T_C\) of \SI{2.94}{\kelvin}. 
Transport measurements at room temperature also showed a deposition temperature dependence of the resistivity (Figure \ref{fig:resist_TiN}). The lowest values of \SI{27}{\micro \ohm\cm} (\SI{47}{\micro\ohm\cm}) for RF deposited \ce{TiN} on \ce{MgO} (DC deposited on \ce{SrTiO3}) was found for deposition temperatures of \SI{830}{\degreeCelsius} (\SI{595}{\degreeCelsius}). These data are in good agreement with the one reported by Shin et al.\cite{Shin:2004tx} The high resistivity value of \SI{833}{\micro \ohm\cm} for \ce{TiN} deposited at \SI{210}{\degreeCelsius} (Figure \ref{fig:resist_TiN}) confirms the assumption that this temperature is too low to achieve crystalline growth of \ce{TiN}. The temperature and substrate dependence of the resistivity was attributed to the different grain sizes and film quality of \ce{TiN} obtained by different sputtering condicions. Due to the lower grain sizes on \ce{SrTiO3} there are more grain boundaries inside the \ce{TiN} films and therefore more perturbations, which leads to a decreased conductance of the layer. Then again, this sample showed the highest transition temperature into the superconducting state, which is an evidence for high film quality with a low amount of impurities inside the layer. On the other hand, we obtained a reverse behavior for samples on \ce{MgO}. The low resistivity of the on \ce{MgO} substrates deposited thin films is attributed to bigger grains and high film quality of the \ce{TiN} compared to samples on \ce{SrTiO3}. In this case, increasing the deposition temperature leads to reduced \(T_C\). This reduction could be caused by oxygen impurities inside the \ce{TiN}.\cite{Muenster:1956} Increasing the deposition temperature causes a diffusion of oxygen atoms from the \ce{MgO} substrate into the \ce{TiN} layer leading to the formation of a \ce{Ti}-\ce{O} interlayer between the substrate and the \ce{TiN}. \ce{Ti}-\ce{O} is a semiconductor and therefore even a small amount inside the thin \ce{TiN} layer disturbes the transition into the superconducting state.

\begin{figure}[t!]%
\includegraphics[width=\linewidth]{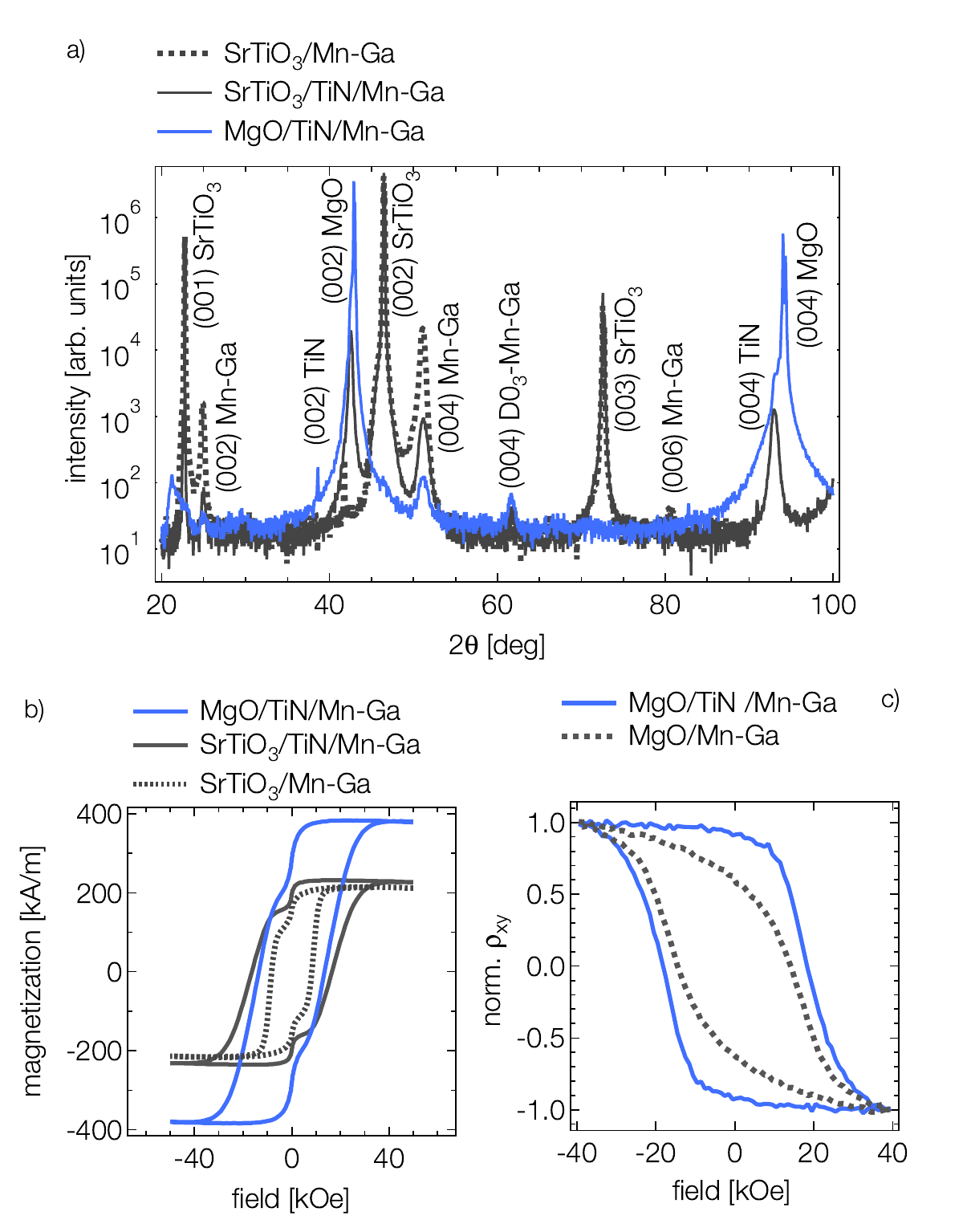}%
\caption{a) XRD pattern of a \ce{Mn_{2.45}Ga} thin film with \ce{TiN} seed layer on \ce{MgO} (blue) and \ce{SrTiO3} (grey) substrate. The deposition temperature was \SI{550}{\degreeCelsius} for the \ce{Mn}-\ce{Ga} thin film and \SI{830}{\degreeCelsius} for the \ce{TiN}. For comparison a plain \ce{Mn_{2.45}Ga} thin film on \ce{SrTiO3} substrate is added (grey dashed), whereas the deposition temperature was \SI{595}{\degreeCelsius}. The \ce{Mn}-\ce{Ga} layer thickness was \SI{25}{\nano\m} for each sample. b)  SQUID and c) AHE hysteresis loops of \ce{Mn_{2.45}Ga} thin films with \ce{TiN} seed layer on \ce{MgO} (blue) and \ce{SrTiO3} (grey) substrate. The deposition temperature of the \ce{Mn}-\ce{Ga} was \SI{550}{\degreeCelsius} and \SI{830}{\degreeCelsius} for the \ce{TiN} buffer. For comparison a plain \ce{Mn_{2.45}Ga} thin film on \ce{SrTiO3} and \ce{MgO} deposited at \SI{595}{\degreeCelsius} is depicted (grey dashed).}%
\label{fig:Mn_Ga_overview}%
\end{figure}%

\subsection{TiN buffered D0\(_{22}\)-Mn-Ga thin films}
To investigate the suitability of \ce{TiN} as a seed layer, we deposited two different Heusler compounds on top of the \ce{TiN}. In a first attempt a perpendicularly magnetized \ce{Mn_{\num{2.45\pm0.06}}Ga} (\SI{25}{\nano\m}) thin film was deposited on a \SI{30}{\nano\m} \ce{TiN} buffer layer on \ce{MgO} and \ce{SrTiO3} substrates. 
The \ce{TiN} seed-layer (\SI{30}{\nano\m} thickness) was DC sputter deposited at deposition temperatures of \SIlist{405;595;830}{\degreeCelsius}, respectively. The Mn-Ga was deposited at \SI{550}{\degreeCelsius} in order to achieve the tetragonally distorted D0\(_{22}\) structure.\cite{Glas:2013vc} For all \ce{TiN} deposition temperatures the \ce{Mn}-\ce{Ga} shows two phases, a mixture of the tetragonal D0\(_{22}\) structure with predicted out-of-plane magnetization and the cubic D0\(_3\) phase with predicted in-plane magnetization (Figure \ref{fig:Mn_Ga_overview} a). However, for \ce{TiN} deposited at \SI{830}{\degreeCelsius} a maximum amount of the D0\(_{22}\) phase was observed. Furthermore, we added a \ce{Mn}-\ce{Ga} thin film on \ce{SrTiO3} substrate without a \ce{TiN} buffer layer for comparison. The optimal deposition temperature to achieve the D0\(_{22}\) structure of the \ce{Mn_{\num{2.45}}Ga} without a buffer layer was \SI{595}{\degreeCelsius} and therefore higher in contrast to \ce{TiN} buffered samples. All samples show the fundamental (004) D0\(_{22}\)-Mn-Ga reflex, corresponding to \(c_{\rm{D0_{22}}} = \SI{7.15}{\angstrom}\) without \ce{TiN} buffer and \SI{7.14}{\angstrom} with \ce{TiN} buffer. The lattice mismatch with the \ce{TiN} buffer leads to an expansion of the \ce{Mn}-\ce{Ga} unit cell along the \(a\)-\(b\) plane, which results in a lower \(c\) lattice constant. Further to the fundamental D0\(_{22}\) peak, the superlattice (002) reflex is visible for all samples. However, the superlattice (006) peak is only visible for \ce{Mn}-\ce{Ga} thin films without \ce{TiN} buffer. This indicates a lower amount and crystallinity of the tetragonally distorted D0\(_{22}\) phase on \ce{TiN} buffered samples. The fundamental (004) peak of the D0\(_{3}\) phase is weakly distinct, corresponding to \(c_{\rm{D0_{3}}} = \SI{5.99}{\angstrom}\) on \ce{SrTiO3} and \(c_{\rm{D0_{3}}} = \SI{6.02}{\angstrom}\) on \ce{MgO}. 
The magnetic properties of \ce{Mn}-\ce{Ga} were investigated using a superconducting quantum interference device. Figure \ref{fig:Mn_Ga_overview} b) depicts the magnetization \(M\) of \ce{TiN} buffered \ce{Mn}-\ce{Ga} thin films on \ce{MgO} (blue) and \ce{SrTiO3} (grey). For comparison an unbuffered \ce{Mn}-\ce{Ga} thin film (grey dashed) is shown. 
An overview of the coercivity \(H\)\(_{c}\) and the squareness \(S_{R}\) of \ce{TiN} buffered \ce{Mn}-\ce{Ga} films on \ce{MgO} and \ce{SrTiO3} is given in Table \ref{tab:Mn-Ga}. We defined the squareness ratio by \(M\)(0\,\rm{kOe})/\(M\)(60\,\rm{kOe}) for the SQUID measurements. \ce{TiN} buffered \ce{Mn_{\num{2.45}}Ga} shows increased coercivity and squareness on both substrate types. To prove this effect, additionally AHE measurements were carried out. Figure \ref{fig:Mn_Ga_overview} c) shows normalized out-of-plane AHE hysteresis curves for \ce{TiN} buffered (blue) and unbuffered \ce{Mn}-\ce{Ga} layers (grey dashed). The \(H\)\(_{c}\) and \(S_{R}\) values are given in Table \ref{tab:Mn-Ga}. In this case we also observe an enhancement of the coercive field and squareness ratio for the \ce{TiN} buffered layer. The squareness ratio is defined by \(\rho_{xy}(0\,\rm{kOe})/\rho_{xy}(40\,\rm{kOe})\). It reveals a stronger enhancement compared to the sample on \ce{SrTiO3} and increases from a value of \(0.62\pm0.05\) to \(0.90\pm0.05\). \ce{TiN} buffered \ce{Mn}-\ce{Ga} thus requires lower deposition temperature compared to unbuffered samples to achieve an out-of-plane magnetization with high coercivity and squareness ratio. The saturation magnetization of \SI{400}{\kilo\ampere/\meter} on \ce{MgO} and \SI{200}{\kilo\ampere/\meter} on \ce{SrTiO3} is in good agreement with the peviously reported results.\cite{Glas:2013vc} Interestingly, the saturation magnetization of the samples grown on \ce{MgO}, is twice as high as for the samples on \ce{SrTiO3}. This behavior was also observed for unbuffered samples and attributed to lower crystallinity due to bigger expansion in the \(a\)-\(b\) plane for \ce{Mn}-\ce{Ga} on \ce{MgO} substrates.\cite{Glas:2013vc} Obviously a \SI{30}{\nano\m} \ce{TiN} layer does not change this behavior. Due to this expansion and therefore the imperfections in the crystal structure, the magnetic moments of the \ce{Mn} atoms occupying the Wyckoff positons \(2b\) and \(4d\) are not compensating each other and this leads to an increase of the magnetization. The observed feature in the SQUID measurements around \num{0}\,\rm{kOe} field (Figure \ref{fig:Mn_Ga_overview} b) is attributed to a second phase (soft magnetic) inside the \ce{Mn}-\ce{Ga} crystal structure, which has a different coercive field. The soft magnetic phase dominates at low field values and therefore leads to a sudden decrease or increase of the magnetization. A similar behavior was already observed for the ternary compounds \ce{Mn}-\ce{Co}-\ce{Ga} and \ce{Mn}-\ce{Fe}-\ce{Ga}.\cite{Fowley:2015gy, Gasi:2013fe}\newline

\subsection{TiN buffered Co\(_{2}\)FeAl thin films}
\begin{figure}[t!]%
\includegraphics[width=\linewidth]{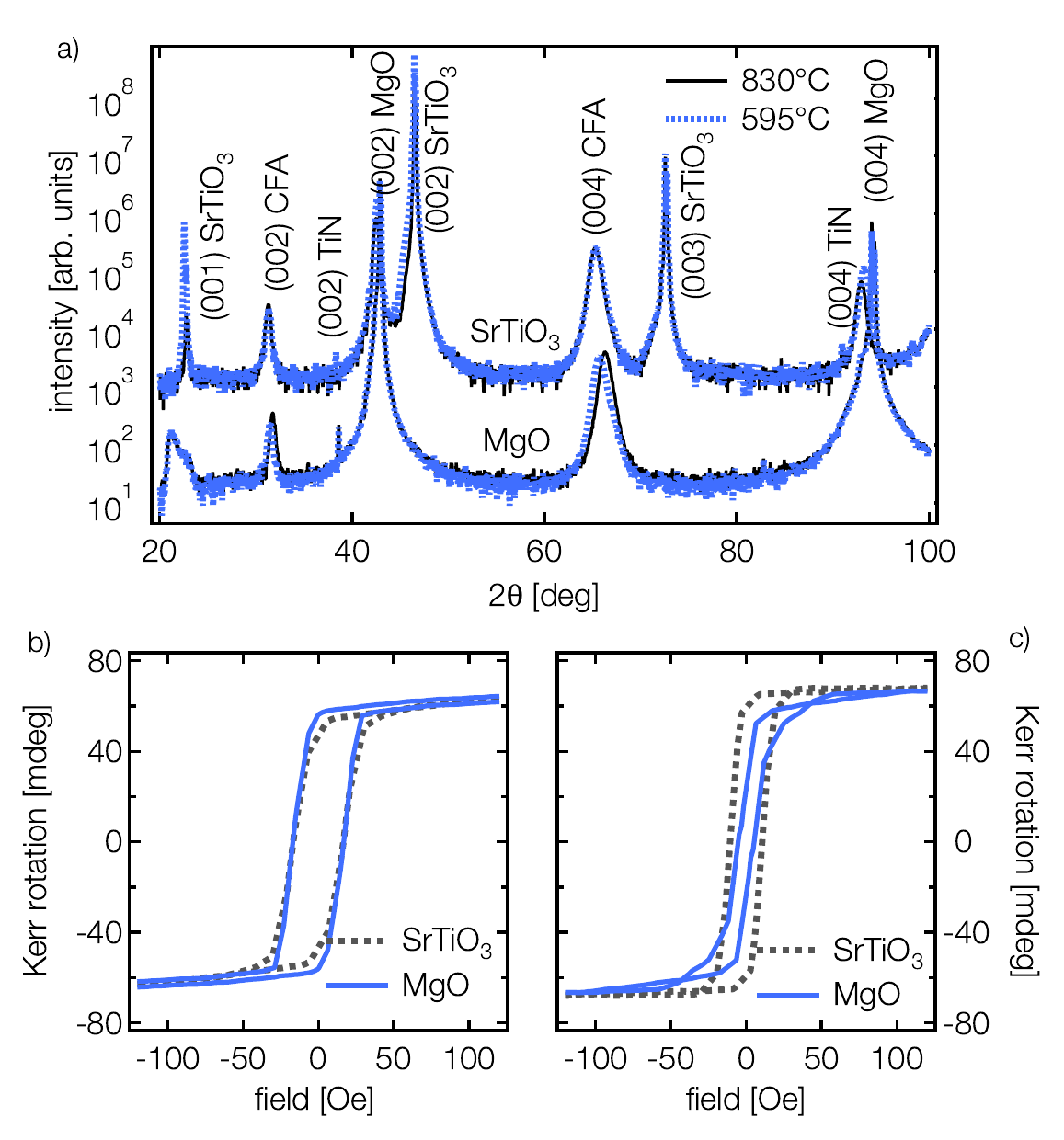}%
\caption{a) XRD pattern of \SI{20}{\nano\m} thick \ce{Co2FeAl} films on \ce{TiN} buffer deposited on \ce{MgO} and \ce{SrTiO3} substrates. The \SI{30}{\nano\m} thick \ce{TiN} layers were deposited at \SIlist{595;830}{\degreeCelsius}, respectively. b) Linear MOKE measurements of \ce{TiN} buffered \ce{Co2FeAl} thin films with a \ce{TiN} deposition temperature of \SI{595}{\degreeCelsius} and c) of \SI{830}{\degreeCelsius} on \ce{MgO} (blue) and \ce{SrTiO3} (grey dashed) substrates.}%
\label{fig:CFA_overview}%
\end{figure}

\begin{table}[!t]
\centering
\begin{tabular}{|c||c||c|}
 \hline
 SQUID measurements&&\\
 \hline
 sample &\(H_{c}\) (kOe) & \(S_{R}\)\\
\hline
\ce{STO}/ \ce{Mn_{2.45}Ga} & \(8 \pm0.5\) & \(0.74 \pm0.05\)\\
\hline
\ce{STO}/\ce{TiN}/\ce{Mn_{2.45}Ga} & \(16 \pm0.5\) & \(0.86 \pm0.05\)\\
\hline
\ce{MgO}/\ce{TiN}/\ce{Mn_{2.45}Ga}& \(13 \pm0.5\) & \(0.75 \pm0.05\)\\
\hline
\hline
AHE measurements&&\\
\hline
sample &\(H_{c}\) (kOe) & \(S_{R}\)\\
\hline
\ce{MgO}/\ce{Mn_{2.45}Ga} & \(13 \pm0.5\) & \(0.62 \pm0.05\)\\
\hline
\ce{MgO}/\ce{TiN}/\ce{Mn_{2.45}Ga} & \(18 \pm0.5\) & \(0.90 \pm0.05\)\\
\hline
\end{tabular}
\caption{Coercive fields and squareness ratios of \ce{TiN} buffered \ce{Mn}-\ce{Ga}. The \ce{TiN} seed-layer (\SI{30}{\nano\m}) was DC sputter deposited at \SI{830}{\degreeCelsius}. \ce{TiN} buffered \ce{Mn}-\ce{Ga} was deposited at \SI{550}{\degreeCelsius}, the unbuffered \ce{Mn}-\ce{Ga} thin film was deposited at \SI{595}{\degreeCelsius}.}
\label{tab:Mn-Ga}
\end{table}
The influence of a TiN buffer layer on the magnetic and crystallographic properties of \ce{Co2FeAl} thin films was investigated in a second step. \SI{20}{\nano\m} and \SI{0.9}{\nano\m} thin \ce{Co2FeAl} layers were sputter deposited on a \SI{30}{\nano\m} thick \ce{TiN} buffer on \ce{MgO} and \ce{SrTiO3} substrates.
The \ce{TiN} was DC sputter deposited at different temperatures (\SIlist{405;595;830}{\degreeCelsius}), whereas the \ce{Co2FeAl} deposition was carried out at room temperature. Figure \ref{fig:CFA_overview} a) shows the obtained XRD spectra. Besides the characteristic \ce{TiN} reflexes, the fundamental and the superlattice peak of the \ce{Co2FeAl} are clearly visible. Even in the as deposited state, the XRD scan reveals a cubic structure with a (002) superlattice peak at 31.3\,degree and a fundamental (004) reflex at 65.3\,degree. Therefore a B2 order of the \ce{Co2FeAl} is proofed with an out-of-plane lattice constant of \num{5.66}\,\si{\angstrom} for \ce{MgO} substrates and \SI{5.72}{\angstrom} for \ce{SrTiO3} substrates. Interestingly, the lattice constant for samples on \ce{SrTiO3} substrates is closer to the literature value of \SI{5.73}{\angstrom}.\cite{BUSCHOW:1983wt} The \ce{TiN} deposition temperature does not influence the crystalline quality of the \ce{Co2FeAl} thin films.

In this case MOKE measurements were carried out to determine the magnetic properties. Figure \ref{fig:CFA_overview} b) and c) illustrates the in-plane hysteresis loops for \ce{TiN} buffered \SI{20}{\nano\m} thick \ce{Co2FeAl} films with a \ce{TiN} deposition temperature of \SIlist{595;830}{\degreeCelsius}. The \ce{Co2FeAl} films reveal sharp switching behavior and in-plane oriented easy magnetization axis, even in the as deposited state. The coercivity and squareness values of \ce{Co2FeAl} on \ce{TiN} are shown in Table \ref{tab:CFA}. The squareness ratio is defined as Kerr rotation at \num{0}\,\rm{Oe} divided by Kerr rotation at \num{150}\,\rm{Oe}. The magnetic properties show a dependence on the substrate type as well as on the \ce{TiN} deposition temperature. With increasing \ce{TiN} deposition temperature, the coercive field decreases. In contrast the squareness ratio for samples on \ce{SrTiO3} increases with increasing \ce{TiN} deposition temperature. As previously mentioned, different deposition temperatures affect the \ce{TiN} seed layer. The temperature during sputtering, and thus the surface energy has a strong influence on the crystal structure and grain size of the material. Structural changes of \ce{TiN} obviously influence the on top deposited material. Especially the grain sizes of the thin \ce{Co2FeAl} layers adjust to the grain structure of the seed layer. The coherence between grain size and coercivity of nanocrystalline ferromagnets was investigated by G. Herzer.\cite{Herzer:1990} He found a \(D^6\) dependence of the coercivity for small grain sizes (up to \SI{50}{\nano\m}) and a \(1/D\) dependence for bigger grains. In our case the grain size of \ce{Co2FeAl} is between \SIlist{25;50}{\nano\m}. Therefore, the decrease of the coercive field of \ce{Co2FeAl} could be explained by a decrease of the grain sizes.

\begin{figure}[t!]%
\includegraphics[width=3in]{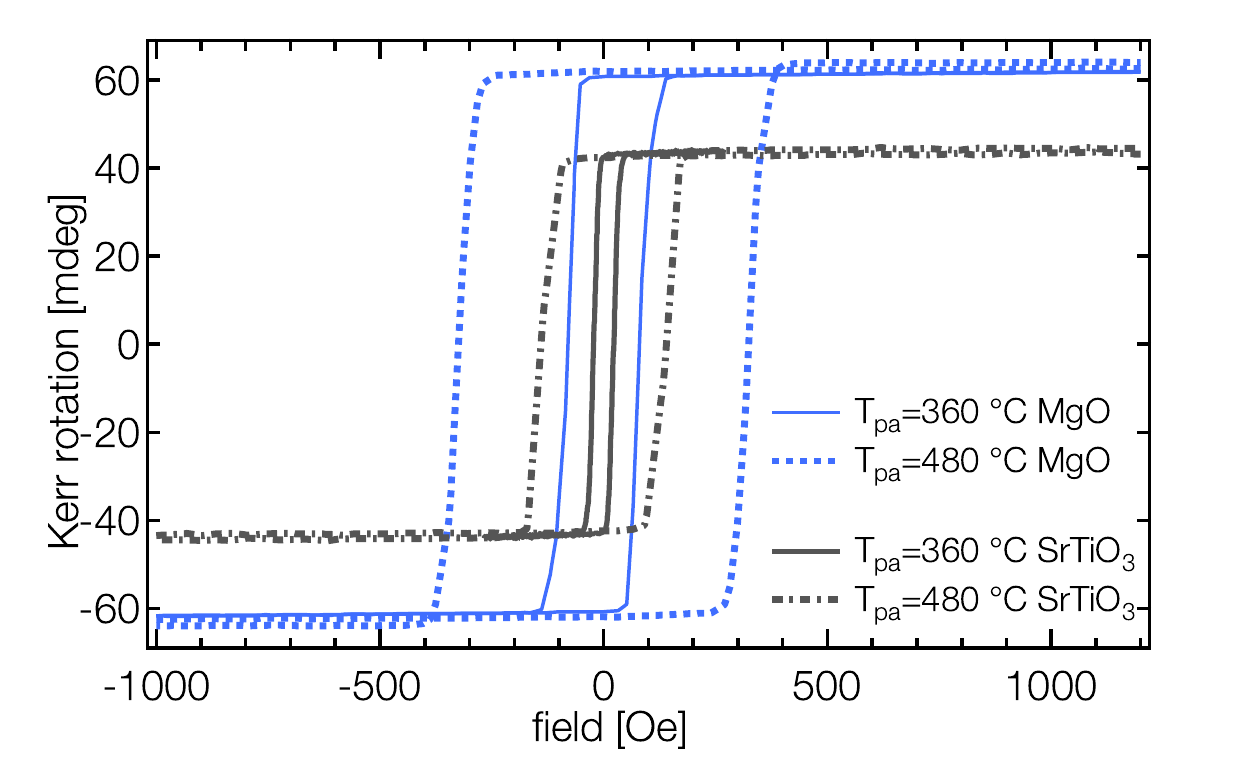}%
\caption{Perpendicular MOKE hysteresis loops of \ce{TiN} (30)/\ce{Co2FeAl} (0.9)/MgO (2) layer systems post annealed at different temperatures (film thicknesses are given in parenthesis). The \ce{TiN} seed-layer was DC sputter deposited at \SI{405}{\degreeCelsius}. High PMA and an increase of the coercive field was observed with increasing post annealing temperature.}%
\label{fig:MOKE_TiN_CFA_pa}%
\end{figure}

\begin{table}[!t]
\centering
\begin{tabular}{|c||c||c||c|}
\hline
\(T_{dep}\) &substrate&\(H_{c}\) (Oe) & \(S_{R}\)\\
\hline
\SI{595}{\degreeCelsius} & \ce{MgO}&\(17 \pm0.5\) & \(0.8 \pm0.05\)\\
\hline
& \ce{STO}& \(17 \pm0.5\) & \(0.65 \pm0.05\)\\
\hline
\SI{830}{\degreeCelsius} & \ce{MgO}&\(7 \pm0.5\) & \(0.2 \pm0.05\)\\
\hline
 & \ce{STO}&\(10 \pm0.5\) & \(0.87 \pm0.05\)\\
\hline
\end{tabular}
\caption{Coercive fields and squareness ratios of \ce{TiN} buffered \ce{Co_{2}FeAl} thin films (\SI{20}{\nano\m}) with ip oriented easy magnetization axis.}
\label{tab:CFA}
\end{table}
\begin{table}[!t]
\centering
\begin{tabular}{|c||c||c|}
\hline
\(T_{pa}\) &\(H_{c}\) (Oe) on \ce{MgO} & \(H_{c}\) (Oe) on \ce{STO}\\
\hline
\SI{360}{\degreeCelsius} &\(76 \pm0.3\) & \(22 \pm0.3\)\\
\hline
\SI{480}{\degreeCelsius} &\(329 \pm0.3\) & \(137 \pm0.1\)\\
\hline
\end{tabular}
\caption{Coercive fields of \ce{TiN} buffered \SI{0.9}{\nano\m} thin \ce{Co_{2}FeAl} layers with oop oriented easy magnetization axis. The squareness ratio is \num{1} for each sample. The \ce{TiN} seed-layer was DC sputter deposited at \SI{405}{\degreeCelsius}.}
\label{tab:CFAoop}
\end{table}
Out-of-plane MOKE measurements for \SI{0.9}{\nano\m} thin, TiN buffered, \ce{Co2FeAl} layers (Figure \ref{fig:MOKE_TiN_CFA_pa}) revealed sharp switching (squareness ratio = \num{1}) and high thermal stability for both substrate types. Even for post annealing temperatures \(T_{pa}\) around \SI{500}{\degreeCelsius} strong perpendicular magnetic anisotropy and an increase of the coercivity was observed. The coercivity of the \ce{Co2FeAl} thin layers post annealed at \SI{360}{\degreeCelsius} is \(76\pm0.3\)\,\rm{Oe} on \ce{MgO} (blue curves) and \(22\pm0.3\)\,\rm{Oe} on \ce{SrTiO3} (grey curves). Post annealing at \SI{480}{\degreeCelsius} leads to a H\(_{c}\) of \(329\pm0.3\)\,\rm{Oe} on \ce{MgO} and \(137\pm0.1\)\,\rm{Oe} on \ce{SrTiO3}. An overview is also given in Table \ref{tab:CFAoop}. Compared to the results of Wen et al.\cite{Wen:2011js} for \ce{Pt} buffered \ce{Co2FeAl} thin films, where the PMA vanishes for post annealing temperatures of \SI{400}{\degreeCelsius}, the \ce {TiN} buffered \ce{Co2FeAl} layers provide high thermal stability, which in turn is beneficial for applications.
\begin{figure}[t!]%
\includegraphics[width=3.7in]{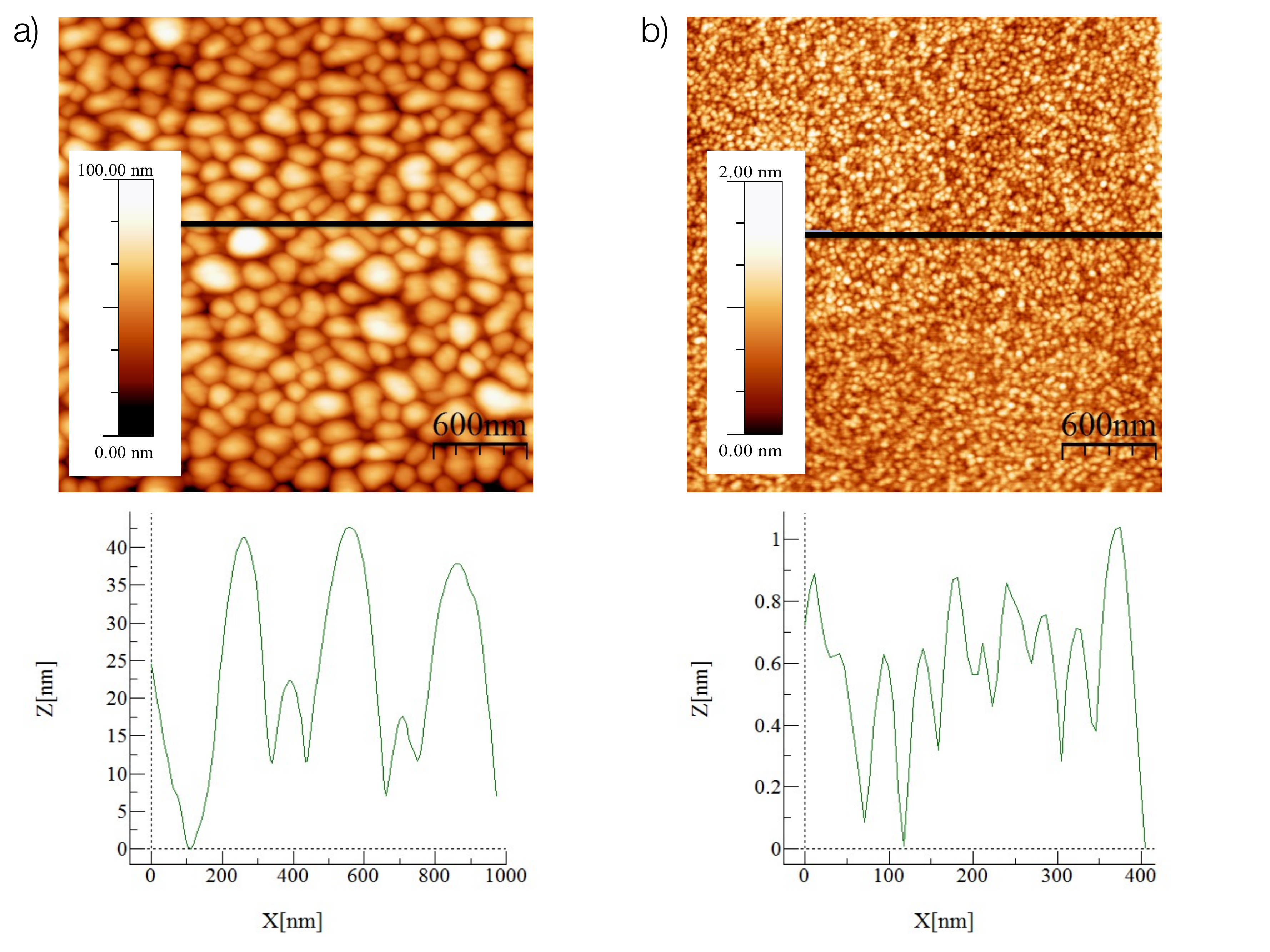}%
\caption{AFM images ($\SI{3}{\micro\m^2}$ section) of TiN buffered a) \ce{Mn_{2.45}Ga} thin film (\SI{25}{\nano\m}) deposited at \SI{550}{\degreeCelsius} and b) \ce{Co2FeAl} thin film (\SI{10}{\nano\m}) deposited at room temperature. The graphs below show the profiles of the sample surface at the marked position (black line).}%
\label{fig:TiN_MnGa_CFA}%
\end{figure}%
Additionally the \ce{Co2FeAl} shows a dependence of the coercivity on the used substrate. We also attribute this behavior to different grain sizes of \ce{TiN} on \ce{MgO} (\ce{SrTiO3}) substrates. As already shown in Figure \ref{fig:AFM_TiN} the grain size of \ce{TiN} on \ce{MgO} is twice as high (\SI{50}{\nano\m}) as on \ce{SrTiO3} (\SI{25}{\nano\m}). For the as deposited state we get a good agreement with a \(D^3\) proportionality between coercivity and grain size. This behavior was explained theoretically by Alben et al. who found this proportionality for materials, where the induced anisotropy \(K_u\) exceeds the structural anisotropy \(K\).\cite{alben1978} In case of \ce{Co2FeAl} this is a necessary condition to obtain the perpendicular magnetocrystalline anisotropy, which we showed by the out-of-plane hysteresis curves. (Figure \ref{fig:MOKE_TiN_CFA_pa}). A detailed discussion of the structural and magnetic properties of \ce{TiN} buffered \ce{Co2FeAl} thin films will be published elsewhere.\newline
Investigations of the surface properties of the two compounds via AFM (Figure \ref{fig:TiN_MnGa_CFA}) revealed strong difference between the surface properties of \ce{TiN} buffered \ce{Mn_{2.45}Ga} and \ce{Co2FeAl}. \ce{Mn_{2.45}Ga} showed high roughness (RMS = \SI{16}{\nano\m}) and island growth (\SI{200}{\nano\m} grain size, determined using profiles of the AFM measurements (Figure \ref{fig:TiN_MnGa_CFA}), which is a strong drawback with regard to the applicability. However, \ce{Co2FeAl} thin films showed smooth surface (RMS = \SI{0.25}{\nano\m}), no island growth and small grains (\SI{50}{\nano\m}). With regard to the integration of this layers into magnetic tunnel junctions, where low roughness plays an important role, \ce{TiN} buffered \ce{Co2FeAl} is a promissing candidate.

\section{Conclusion}
We successfully prepared (001) oriented \ce{TiN} thin films on \ce{MgO} and \ce{SrTiO3} substrates. XAS measurements revealed the formation of pure \ce{TiN} in the thin films. A transition into the superconducting state was observed below \SI{5}{\kelvin}. Transport measurements at room temperature showed a resistivity of \SIlist{27;47}{\micro\ohm \centi\meter} on \ce{MgO} and \ce{SrTiO3} substrates, respectively. It has been demonstrated that out-of-plane magnetized \ce{Mn_{2.45}Ga}  and \ce{Co2FeAl} thin films crystallize well on a \ce{TiN} seed layer. Even in the as deposited state \ce{Co2FeAl} provides the B2 crystal structure. \ce{Mn}-\ce{Ga} thin films exhibit higher coercivity and squareness ratio when prepared on a \ce{TiN} buffer. In addition a lower deposition temperature for \ce{TiN} buffered systems was observed. In-plane magnetized \SI{20}{\nano\m} thick \ce{Co2FeAl} films revealed high squareness ratio even in the as prepared state. \SI{0.9}{\nano\m} thin \ce{Co2FeAl} with out-of-plane oriented easy magnetization axis showed high thermal stability for temperatures up to \SI{500}{\degreeCelsius}.
\ce{TiN} provides various advantages. The low resistivity makes \ce{TiN} a promising seed layer for Heusler compounds in MTJ's. Besides the low roughness, high thermal stability and conductivity, it also enhances the out-of-plane magnetocrystalline anisotropy and optimizes the switching behavior of the used Heusler material. In case of \ce{Co2FeAl}, \ce{TiN} additionally enhances the thermal stability and therefore is highly preferable for applications.

\section*{Acknowledgments}
The authors gratefully acknowledge financial support by the Deutsche Forschungsgemeinschaft (DFG, Contract No. RE 1052/32-1) and the opportunity to work at BL 6.3.1 of the Advanced Light Source, Berkeley, USA, which is supported by the Director, Office of Science, Office of Basic Energy Sciences, of the U.S. Department of Energy under Contract No. DE-AC02-05CH11231. D.E. is financially supported by the ERC Advanced Grant (291472 Idea Heusler).

\bibliographystyle{apsrev4-1}
\bibliography{bibfile}

\begin{thebibliography}{19}%
\makeatletter
\providecommand \@ifxundefined [1]{%
 \@ifx{#1\undefined}
}%
\providecommand \@ifnum [1]{%
 \ifnum #1\expandafter \@firstoftwo
 \else \expandafter \@secondoftwo
 \fi
}%
\providecommand \@ifx [1]{%
 \ifx #1\expandafter \@firstoftwo
 \else \expandafter \@secondoftwo
 \fi
}%
\providecommand \natexlab [1]{#1}%
\providecommand \enquote  [1]{``#1''}%
\providecommand \bibnamefont  [1]{#1}%
\providecommand \bibfnamefont [1]{#1}%
\providecommand \citenamefont [1]{#1}%
\providecommand \href@noop [0]{\@secondoftwo}%
\providecommand \href [0]{\begingroup \@sanitize@url \@href}%
\providecommand \@href[1]{\@@startlink{#1}\@@href}%
\providecommand \@@href[1]{\endgroup#1\@@endlink}%
\providecommand \@sanitize@url [0]{\catcode `\\12\catcode `\$12\catcode
  `\&12\catcode `\#12\catcode `\^12\catcode `\_12\catcode `\%12\relax}%
\providecommand \@@startlink[1]{}%
\providecommand \@@endlink[0]{}%
\providecommand \url  [0]{\begingroup\@sanitize@url \@url }%
\providecommand \@url [1]{\endgroup\@href {#1}{\urlprefix }}%
\providecommand \urlprefix  [0]{URL }%
\providecommand \Eprint [0]{\href }%
\providecommand \doibase [0]{http://dx.doi.org/}%
\providecommand \selectlanguage [0]{\@gobble}%
\providecommand \bibinfo  [0]{\@secondoftwo}%
\providecommand \bibfield  [0]{\@secondoftwo}%
\providecommand \translation [1]{[#1]}%
\providecommand \BibitemOpen [0]{}%
\providecommand \bibitemStop [0]{}%
\providecommand \bibitemNoStop [0]{.\EOS\space}%
\providecommand \EOS [0]{\spacefactor3000\relax}%
\providecommand \BibitemShut  [1]{\csname bibitem#1\endcsname}%
\let\auto@bib@innerbib\@empty
\bibitem [{\citenamefont {Versluijs}\ \emph {et~al.}(2001)\citenamefont
  {Versluijs}, \citenamefont {Bari},\ and\ \citenamefont
  {Coey}}]{Versluijs:2001uf}%
  \BibitemOpen
  \bibfield  {author} {\bibinfo {author} {\bibfnamefont {J.~J.}\ \bibnamefont
  {Versluijs}}, \bibinfo {author} {\bibfnamefont {M.~A.}\ \bibnamefont {Bari}},
  \ and\ \bibinfo {author} {\bibfnamefont {J.}~\bibnamefont {Coey}},\
  }\href@noop {} {\bibfield  {journal} {\bibinfo  {journal} {Physical Review
  Letters}\ }\textbf {\bibinfo {volume} {87}},\ \bibinfo {pages} {026601}
  (\bibinfo {year} {2001})}\BibitemShut {NoStop}%
\bibitem [{\citenamefont {Degroot}\ \emph {et~al.}(1983)\citenamefont
  {Degroot}, \citenamefont {Mueller}, \citenamefont {Vanengen},\ and\
  \citenamefont {Buschow}}]{DEGROOT:1983vn}%
  \BibitemOpen
  \bibfield  {author} {\bibinfo {author} {\bibfnamefont {R.}~\bibnamefont
  {Degroot}}, \bibinfo {author} {\bibfnamefont {F.}~\bibnamefont {Mueller}},
  \bibinfo {author} {\bibfnamefont {P.}~\bibnamefont {Vanengen}}, \ and\
  \bibinfo {author} {\bibfnamefont {K.}~\bibnamefont {Buschow}},\ }\href@noop
  {} {\bibfield  {journal} {\bibinfo  {journal} {Physical Review Letters}\
  }\textbf {\bibinfo {volume} {50}},\ \bibinfo {pages} {2024} (\bibinfo {year}
  {1983})}\BibitemShut {NoStop}%
\bibitem [{\citenamefont {Sterwerf}\ \emph {et~al.}(2013)\citenamefont
  {Sterwerf}, \citenamefont {Meinert}, \citenamefont {Schmalhorst},\ and\
  \citenamefont {Reiss}}]{Sterwerf:2013it}%
  \BibitemOpen
  \bibfield  {author} {\bibinfo {author} {\bibfnamefont {C.}~\bibnamefont
  {Sterwerf}}, \bibinfo {author} {\bibfnamefont {M.}~\bibnamefont {Meinert}},
  \bibinfo {author} {\bibfnamefont {J.-M.}\ \bibnamefont {Schmalhorst}}, \ and\
  \bibinfo {author} {\bibfnamefont {G.}~\bibnamefont {Reiss}},\ }\href@noop {}
  {\bibfield  {journal} {\bibinfo  {journal} {IEEE Transactions on Magnetics}\
  }\textbf {\bibinfo {volume} {49}},\ \bibinfo {pages} {4386} (\bibinfo {year}
  {2013})}\BibitemShut {NoStop}%
\bibitem [{\citenamefont {Balke}\ \emph {et~al.}(2007)\citenamefont {Balke},
  \citenamefont {Fecher}, \citenamefont {Winterlik},\ and\ \citenamefont
  {Felser}}]{Balke:2007eb}%
  \BibitemOpen
  \bibfield  {author} {\bibinfo {author} {\bibfnamefont {B.}~\bibnamefont
  {Balke}}, \bibinfo {author} {\bibfnamefont {G.~H.}\ \bibnamefont {Fecher}},
  \bibinfo {author} {\bibfnamefont {J.}~\bibnamefont {Winterlik}}, \ and\
  \bibinfo {author} {\bibfnamefont {C.}~\bibnamefont {Felser}},\ }\href@noop {}
  {\bibfield  {journal} {\bibinfo  {journal} {Applied Physics Letters}\
  }\textbf {\bibinfo {volume} {90}},\ \bibinfo {pages} {152504} (\bibinfo
  {year} {2007})}\BibitemShut {NoStop}%
\bibitem [{\citenamefont {Glas}\ \emph
  {et~al.}(2013{\natexlab{a}})\citenamefont {Glas}, \citenamefont {Sterwerf},
  \citenamefont {Schmalhorst}, \citenamefont {Ebke}, \citenamefont {Jenkins},
  \citenamefont {Arenholz},\ and\ \citenamefont {Reiss}}]{Glas:2013we}%
  \BibitemOpen
  \bibfield  {author} {\bibinfo {author} {\bibfnamefont {M.}~\bibnamefont
  {Glas}}, \bibinfo {author} {\bibfnamefont {C.}~\bibnamefont {Sterwerf}},
  \bibinfo {author} {\bibfnamefont {J.-M.}\ \bibnamefont {Schmalhorst}},
  \bibinfo {author} {\bibfnamefont {D.}~\bibnamefont {Ebke}}, \bibinfo {author}
  {\bibfnamefont {C.}~\bibnamefont {Jenkins}}, \bibinfo {author} {\bibfnamefont
  {E.}~\bibnamefont {Arenholz}}, \ and\ \bibinfo {author} {\bibfnamefont
  {G.}~\bibnamefont {Reiss}},\ }\href@noop {} {\bibfield  {journal} {\bibinfo
  {journal} {Journal of Applied Physics}\ }\textbf {\bibinfo {volume} {114}},\
  \bibinfo {pages} {183910} (\bibinfo {year} {2013}{\natexlab{a}})}\BibitemShut
  {NoStop}%
\bibitem [{\citenamefont {Ebke}\ \emph {et~al.}(2010)\citenamefont {Ebke},
  \citenamefont {Thomas}, \citenamefont {Schebaum}, \citenamefont
  {Sch{\"a}fers}, \citenamefont {Nissen}, \citenamefont {Drewello},
  \citenamefont {H{\"u}tten},\ and\ \citenamefont {Thomas}}]{Ebke:2010ci}%
  \BibitemOpen
  \bibfield  {author} {\bibinfo {author} {\bibfnamefont {D.}~\bibnamefont
  {Ebke}}, \bibinfo {author} {\bibfnamefont {P.}~\bibnamefont {Thomas}},
  \bibinfo {author} {\bibfnamefont {O.}~\bibnamefont {Schebaum}}, \bibinfo
  {author} {\bibfnamefont {M.}~\bibnamefont {Sch{\"a}fers}}, \bibinfo {author}
  {\bibfnamefont {D.}~\bibnamefont {Nissen}}, \bibinfo {author} {\bibfnamefont
  {V.}~\bibnamefont {Drewello}}, \bibinfo {author} {\bibfnamefont
  {A.}~\bibnamefont {H{\"u}tten}}, \ and\ \bibinfo {author} {\bibfnamefont
  {A.}~\bibnamefont {Thomas}},\ }\href@noop {} {\bibfield  {journal} {\bibinfo
  {journal} {Journal of Magnetism and Magnetic Materials}\ }\textbf {\bibinfo
  {volume} {322}},\ \bibinfo {pages} {996} (\bibinfo {year}
  {2010})}\BibitemShut {NoStop}%
\bibitem [{\citenamefont {Pritschow}(2007)}]{Pritschow:07}%
  \BibitemOpen
  \bibfield  {author} {\bibinfo {author} {\bibfnamefont {M.}~\bibnamefont
  {Pritschow}},\ }\href@noop {} {Ph.D. thesis},\ \bibinfo  {school} {Institut
  f{\"u}r Mikroelektronik Stuttgart} (\bibinfo {year} {2007})\BibitemShut
  {NoStop}%
\bibitem [{\citenamefont {Magnus}\ \emph {et~al.}(2011)\citenamefont {Magnus},
  \citenamefont {Ingason}, \citenamefont {Olafsson},\ and\ \citenamefont
  {Gudmundsson}}]{Magnus:2011wl}%
  \BibitemOpen
  \bibfield  {author} {\bibinfo {author} {\bibfnamefont {F.}~\bibnamefont
  {Magnus}}, \bibinfo {author} {\bibfnamefont {A.~S.}\ \bibnamefont {Ingason}},
  \bibinfo {author} {\bibfnamefont {S.}~\bibnamefont {Olafsson}}, \ and\
  \bibinfo {author} {\bibfnamefont {J.~T.}\ \bibnamefont {Gudmundsson}},\
  }\href@noop {} {\bibfield  {journal} {\bibinfo  {journal} {Thin Solid Films}\
  }\textbf {\bibinfo {volume} {520(5)}},\ \bibinfo {pages} {1621} (\bibinfo
  {year} {2011})}\BibitemShut {NoStop}%
\bibitem [{\citenamefont {Krockenberger}\ \emph {et~al.}(2012)\citenamefont
  {Krockenberger}, \citenamefont {Karimoto}, \citenamefont {Yamamoto},\ and\
  \citenamefont {Semba}}]{Krockenberger:2012bo}%
  \BibitemOpen
  \bibfield  {author} {\bibinfo {author} {\bibfnamefont {Y.}~\bibnamefont
  {Krockenberger}}, \bibinfo {author} {\bibfnamefont {S.-i.}\ \bibnamefont
  {Karimoto}}, \bibinfo {author} {\bibfnamefont {H.}~\bibnamefont {Yamamoto}},
  \ and\ \bibinfo {author} {\bibfnamefont {K.}~\bibnamefont {Semba}},\
  }\href@noop {} {\bibfield  {journal} {\bibinfo  {journal} {Journal of Applied
  Physics}\ }\textbf {\bibinfo {volume} {112}},\ \bibinfo {pages} {083920}
  (\bibinfo {year} {2012})}\BibitemShut {NoStop}%
\bibitem [{\citenamefont {Soriano}\ \emph {et~al.}(1993)\citenamefont
  {Soriano}, \citenamefont {Abbate}, \citenamefont {Pen}, \citenamefont
  {Czy{\.z}yk},\ and\ \citenamefont {Fuggle}}]{Soriano:1993vd}%
  \BibitemOpen
  \bibfield  {author} {\bibinfo {author} {\bibfnamefont {L.}~\bibnamefont
  {Soriano}}, \bibinfo {author} {\bibfnamefont {M.}~\bibnamefont {Abbate}},
  \bibinfo {author} {\bibfnamefont {H.}~\bibnamefont {Pen}}, \bibinfo {author}
  {\bibfnamefont {M.~T.}\ \bibnamefont {Czy{\.z}yk}}, \ and\ \bibinfo {author}
  {\bibfnamefont {J.~C.}\ \bibnamefont {Fuggle}},\ }\href@noop {} {\bibfield
  {journal} {\bibinfo  {journal} {Journal of electron spectroscopy and related
  phenomena}\ }\textbf {\bibinfo {volume} {62}},\ \bibinfo {pages} {197}
  (\bibinfo {year} {1993})}\BibitemShut {NoStop}%
\bibitem [{\citenamefont {Shin}\ \emph {et~al.}(2004)\citenamefont {Shin},
  \citenamefont {Rudenja}, \citenamefont {Gall}, \citenamefont {Hellgren},
  \citenamefont {Lee}, \citenamefont {Petrov},\ and\ \citenamefont
  {Greene}}]{Shin:2004tx}%
  \BibitemOpen
  \bibfield  {author} {\bibinfo {author} {\bibfnamefont {C.-S.}\ \bibnamefont
  {Shin}}, \bibinfo {author} {\bibfnamefont {S.}~\bibnamefont {Rudenja}},
  \bibinfo {author} {\bibfnamefont {D.}~\bibnamefont {Gall}}, \bibinfo {author}
  {\bibfnamefont {N.}~\bibnamefont {Hellgren}}, \bibinfo {author}
  {\bibfnamefont {T.-Y.}\ \bibnamefont {Lee}}, \bibinfo {author} {\bibfnamefont
  {I.}~\bibnamefont {Petrov}}, \ and\ \bibinfo {author} {\bibfnamefont {J.~E.}\
  \bibnamefont {Greene}},\ }\href@noop {} {\bibfield  {journal} {\bibinfo
  {journal} {Journal of Applied Physics}\ }\textbf {\bibinfo {volume} {95}},\
  \bibinfo {pages} {356} (\bibinfo {year} {2004})}\BibitemShut {NoStop}%
\bibitem [{\citenamefont {M{\"u}nster}\ and\ \citenamefont
  {Sagel}(1956)}]{Muenster:1956}%
  \BibitemOpen
  \bibfield  {author} {\bibinfo {author} {\bibfnamefont {A.}~\bibnamefont
  {M{\"u}nster}}\ and\ \bibinfo {author} {\bibfnamefont {K.}~\bibnamefont
  {Sagel}},\ }\href@noop {} {\bibfield  {journal} {\bibinfo  {journal}
  {Zeitschrift f{\"u}r Physik}\ ,\ \bibinfo {pages} {139}} (\bibinfo {year}
  {1956})}\BibitemShut {NoStop}%
\bibitem [{\citenamefont {Glas}\ \emph
  {et~al.}(2013{\natexlab{b}})\citenamefont {Glas}, \citenamefont {Ebke},
  \citenamefont {Imort}, \citenamefont {Thomas},\ and\ \citenamefont
  {Reiss}}]{Glas:2013vc}%
  \BibitemOpen
  \bibfield  {author} {\bibinfo {author} {\bibfnamefont {M.}~\bibnamefont
  {Glas}}, \bibinfo {author} {\bibfnamefont {D.}~\bibnamefont {Ebke}}, \bibinfo
  {author} {\bibfnamefont {I.~M.}\ \bibnamefont {Imort}}, \bibinfo {author}
  {\bibfnamefont {P.}~\bibnamefont {Thomas}}, \ and\ \bibinfo {author}
  {\bibfnamefont {G.}~\bibnamefont {Reiss}},\ }\href@noop {} {\bibfield
  {journal} {\bibinfo  {journal} {Journal of Magnetism and Magnetic Materials}\
  }\textbf {\bibinfo {volume} {333}},\ \bibinfo {pages} {134} (\bibinfo {year}
  {2013}{\natexlab{b}})}\BibitemShut {NoStop}%
\bibitem [{\citenamefont {Fowley}\ \emph {et~al.}(2015)\citenamefont {Fowley},
  \citenamefont {Ouardi}, \citenamefont {Kubota}, \citenamefont {Yildirim},
  \citenamefont {Neudert}, \citenamefont {Lenz}, \citenamefont {Sluka},
  \citenamefont {Lindner}, \citenamefont {Law}, \citenamefont {Mizukami},
  \citenamefont {Fecher}, \citenamefont {Felser},\ and\ \citenamefont
  {Deac}}]{Fowley:2015gy}%
  \BibitemOpen
  \bibfield  {author} {\bibinfo {author} {\bibfnamefont {C.}~\bibnamefont
  {Fowley}}, \bibinfo {author} {\bibfnamefont {S.}~\bibnamefont {Ouardi}},
  \bibinfo {author} {\bibfnamefont {T.}~\bibnamefont {Kubota}}, \bibinfo
  {author} {\bibfnamefont {O.}~\bibnamefont {Yildirim}}, \bibinfo {author}
  {\bibfnamefont {A.}~\bibnamefont {Neudert}}, \bibinfo {author} {\bibfnamefont
  {K.}~\bibnamefont {Lenz}}, \bibinfo {author} {\bibfnamefont {V.}~\bibnamefont
  {Sluka}}, \bibinfo {author} {\bibfnamefont {J.}~\bibnamefont {Lindner}},
  \bibinfo {author} {\bibfnamefont {J.~M.}\ \bibnamefont {Law}}, \bibinfo
  {author} {\bibfnamefont {S.}~\bibnamefont {Mizukami}}, \bibinfo {author}
  {\bibfnamefont {G.~H.}\ \bibnamefont {Fecher}}, \bibinfo {author}
  {\bibfnamefont {C.}~\bibnamefont {Felser}}, \ and\ \bibinfo {author}
  {\bibfnamefont {A.~M.}\ \bibnamefont {Deac}},\ }\href@noop {} {\bibfield
  {journal} {\bibinfo  {journal} {Journal of Physics D: Applied Physics}\
  }\textbf {\bibinfo {volume} {48}},\ \bibinfo {pages} {164006} (\bibinfo
  {year} {2015})}\BibitemShut {NoStop}%
\bibitem [{\citenamefont {Gasi}\ \emph {et~al.}(2013)\citenamefont {Gasi},
  \citenamefont {Nayak}, \citenamefont {Winterlik}, \citenamefont
  {Ksenofontov}, \citenamefont {Adler}, \citenamefont {Nicklas},\ and\
  \citenamefont {Felser}}]{Gasi:2013fe}%
  \BibitemOpen
  \bibfield  {author} {\bibinfo {author} {\bibfnamefont {T.}~\bibnamefont
  {Gasi}}, \bibinfo {author} {\bibfnamefont {A.~K.}\ \bibnamefont {Nayak}},
  \bibinfo {author} {\bibfnamefont {J.}~\bibnamefont {Winterlik}}, \bibinfo
  {author} {\bibfnamefont {V.}~\bibnamefont {Ksenofontov}}, \bibinfo {author}
  {\bibfnamefont {P.}~\bibnamefont {Adler}}, \bibinfo {author} {\bibfnamefont
  {M.}~\bibnamefont {Nicklas}}, \ and\ \bibinfo {author} {\bibfnamefont
  {C.}~\bibnamefont {Felser}},\ }\href@noop {} {\bibfield  {journal} {\bibinfo
  {journal} {Applied Physics Letters}\ }\textbf {\bibinfo {volume} {102}},\
  \bibinfo {pages} {202402} (\bibinfo {year} {2013})}\BibitemShut {NoStop}%
\bibitem [{\citenamefont {Buschow}\ \emph {et~al.}(1983)\citenamefont
  {Buschow}, \citenamefont {Van~Engen},\ and\ \citenamefont
  {Jongebreur}}]{BUSCHOW:1983wt}%
  \BibitemOpen
  \bibfield  {author} {\bibinfo {author} {\bibfnamefont {K.}~\bibnamefont
  {Buschow}}, \bibinfo {author} {\bibfnamefont {P.~G.}\ \bibnamefont
  {Van~Engen}}, \ and\ \bibinfo {author} {\bibfnamefont {R.}~\bibnamefont
  {Jongebreur}},\ }\href@noop {} {\bibfield  {journal} {\bibinfo  {journal}
  {Journal of Magnetism and Magnetic Materials}\ }\textbf {\bibinfo {volume}
  {38}},\ \bibinfo {pages} {1} (\bibinfo {year} {1983})}\BibitemShut {NoStop}%
\bibitem [{\citenamefont {Herzer}(1990)}]{Herzer:1990}%
  \BibitemOpen
  \bibfield  {author} {\bibinfo {author} {\bibfnamefont {G.}~\bibnamefont
  {Herzer}},\ }\href@noop {} {\bibfield  {journal} {\bibinfo  {journal} {IEEE
  Transactions on Magnetics}\ }\textbf {\bibinfo {volume} {26}},\ \bibinfo
  {pages} {1397} (\bibinfo {year} {1990})}\BibitemShut {NoStop}%
\bibitem [{\citenamefont {Wen}\ \emph {et~al.}(2011)\citenamefont {Wen},
  \citenamefont {Sukegawa}, \citenamefont {Mitani},\ and\ \citenamefont
  {Inomata}}]{Wen:2011js}%
  \BibitemOpen
  \bibfield  {author} {\bibinfo {author} {\bibfnamefont {Z.}~\bibnamefont
  {Wen}}, \bibinfo {author} {\bibfnamefont {H.}~\bibnamefont {Sukegawa}},
  \bibinfo {author} {\bibfnamefont {S.}~\bibnamefont {Mitani}}, \ and\ \bibinfo
  {author} {\bibfnamefont {K.}~\bibnamefont {Inomata}},\ }\href@noop {}
  {\bibfield  {journal} {\bibinfo  {journal} {Applied Physics Letters}\
  }\textbf {\bibinfo {volume} {98}},\ \bibinfo {pages} {192505} (\bibinfo
  {year} {2011})}\BibitemShut {NoStop}%
\bibitem [{\citenamefont {Alben}\ \emph {et~al.}(1978)\citenamefont {Alben},
  \citenamefont {Becker},\ and\ \citenamefont {Chi}}]{alben1978}%
  \BibitemOpen
  \bibfield  {author} {\bibinfo {author} {\bibfnamefont {R.}~\bibnamefont
  {Alben}}, \bibinfo {author} {\bibfnamefont {J.}~\bibnamefont {Becker}}, \
  and\ \bibinfo {author} {\bibfnamefont {M.}~\bibnamefont {Chi}},\ }\href@noop
  {} {\bibfield  {journal} {\bibinfo  {journal} {Journal of Applied Physics}\
  }\textbf {\bibinfo {volume} {49}},\ \bibinfo {pages} {1653} (\bibinfo {year}
  {1978})}\BibitemShut {NoStop}%
\end{thebibliography}%

\end{document}